\documentclass[paper]{JHEP3}

\usepackage{epsfig,multicol,multirow}
\usepackage{amsmath,subfigure,latexsym,amssymb}
\usepackage{here}

\def\lsi{\raise0.3ex\hbox{$<$\kern-0.75em\raise-1.1ex\hbox{$\sim$}}}
\def\gsi{\raise0.3ex\hbox{$>$\kern-0.75em\raise-1.1ex\hbox{$\sim$}}}

\newcommand\fverb{\setbox\pippobox=\hbox\bgroup\verb}
\newcommand\fverbdo{\egroup\medskip\noindent%
                        \fbox{\unhbox\pippobox}\ }
\newcommand\fverbit{\egroup\item[\fbox{\unhbox\pippobox}]}
\newbox\pippobox
\newcommand{\beq}{\begin{equation}}
\newcommand{\eeq}{\end{equation}}
\newcommand{\beqa}{\begin{eqnarray}}
\newcommand{\eeqa}{\end{eqnarray}}
\newcommand{\n}{\nonumber \\}

\newcommand{\id}{{1\!\!1}} 
\def\e{{\,\rm e}\,}

\newcommand{\del}{\partial}
\newcommand {\tr}{{\rm tr\,}}

\newcommand {\CTr}{{\cal T} \! r\,}

\preprint{
SAGA-HE-253 \\ 
KEK-TH-1321
}
\title{
Dominance of a single topological sector\\
in gauge theory on non-commutative geometry
}

\author{
Hajime Aoki,${}^{a}$
Jun Nishimura${}^{b,c}$ and
Yoshiaki Susaki${}^{b}$\\ 
\llap{$^a$}Department of Physics, Saga University, 
Saga 840-8502, Japan \\
\llap{$^b$}KEK Theory Center, High Energy Accelerator Research Organization, \\
Tsukuba, Ibaraki, 305-0801, Japan \\
\llap{$^c$}Department of Particle and Nuclear Physics,\\
Graduate University for Advanced Studies (SOKENDAI),\\
Tsukuba, Ibaraki 305-0801, Japan \\
\email{haoki@cc.saga-u.ac.jp,
jnishi@post.kek.jp,
susaki@post.kek.jp}} 

\abstract{
We demonstrate a striking effect
of non-commutative (NC) geometry on
topological properties of gauge theory
by Monte Carlo simulations.
We study 2d U(1) NC gauge theory 
for various boundary conditions
using a new finite-matrix formulation proposed recently.
We find that a single topological sector
dictated by the boundary condition
dominates in the continuum limit.
This is in sharp contrast to the results in 
commutative space-time based on lattice gauge theory, 
where all topological sectors appear with certain weights
in the continuum limit.
We discuss
possible implications of this effect
in the context of
string theory compactifications
and in field theory contexts.
}


\begin{document}

\section{Introduction}
\label{sec:intro}

In matrix models of superstring 
theories \cite{Banks:1996vh,Ishibashi:1996xs},
space and possibly also time are described by matrices,
which are non-commutative from an outset.
This matches well with
the idea of non-commutative (NC) geometry \cite{Sny,Connes},
which 
has been discussed for many years
as possible effects of quantum gravity \cite{gravity}.
The connection became more concrete
since field theories on a NC geometry
were shown to appear naturally from matrix models \cite{CDS,AIIKKT}
and from string theories \cite{String}\footnote{Recently the idea of
emergent gravity is studied intensively \cite{Steinacker:2007dq}
as another connection between quantum gravity
and NC
geometry.}. 
Dynamical properties of such NC field theories 
are therefore expected to 
play an important role 
in understanding fundamental nature of our real world.

In this paper, we demonstrate a striking effect
of NC geometry on topological properties of gauge theory.
We perform Monte Carlo simulations of
2d U(1) gauge theory with various boundary conditions
using a new finite-matrix formulation 
proposed recently \cite{Aoki:2008ik}.
The gauge field configurations are classified
into topological sectors by the topological charge or
the index of the Dirac operator.
We find that a single topological sector dictated
by the boundary condition dominates in the continuum limit.
This is in sharp contrast to the results 
in commutative space-time
based on lattice gauge theory \cite{Hassan:1995dn,GHL}.
There, the distribution of the topological charge is gaussian 
with a finite extent.
This means, in particular, that 
all topological sectors appear
from a theory with a specific (e.g.,\ periodic) 
boundary condition.
Moreover, the width of the distribution
diverges 
as the physical volume is increased,
meaning that every topological sector appears with an equal weight
in the thermodynamic limit.

We may interpret this striking difference as a kind
of smoothing effects of NC geometry.
Note that in both commutative and NC space-times, a 
regular configuration,
which can be made smooth by an appropriate gauge transformation,
has a specific 
topological charge
fixed by the boundary condition.
In commutative space-time, one can construct non-regular
configurations
without increasing the action considerably.
This makes it possible to obtain a configuration in different
topological sectors rather easily. 
In NC space-time, on the other hand, such non-regular configurations
increase the action significantly since the star-product contains
all higher derivative terms.
We confirm this point of view 
by calculating the average
action for each topological sector.
Indeed we observe that the average action is smaller
in the specific topological sector 
than in the other sectors.

Thus, in NC gauge theory, one topological sector is singled out
depending on the boundary condition.
This may 
be important in the context of 
string theory compactifications, in which the topological properties
in the extra dimensions determine, for instance, 
the number of generations.
If the space-time in the extra dimensions are actually non-commutative
(See refs.\ \cite{Aschieri:2003vy, AIMN}
for discussions on models, in which extra dimensions appear as a fuzzy 
sphere.),
our results suggest a mechanism for singling out a particular
number of generations etc..
We also discuss
possible implications on
problems related to topological aspects of field theory 
such as the baryon number asymmetry and the strong CP problem.


The rest of this paper is organized as follows.
In section \ref{sec:formulation}
we briefly review the finite-matrix formulation
of gauge theories on a NC torus,
which has been generalized recently to
allow for twisted boundary conditions.
In section \ref{sec:index}
we present our Monte Carlo results for the distribution
of topological sectors.
Section \ref{sec:summary} is devoted
to a summary and discussions.

\section{Brief review of the finite-matrix formulation}
\label{sec:formulation}

In this section we briefly review the finite-matrix formulation 
of gauge theories on a NC torus with twisted boundary conditions 
\cite{Aoki:2008ik}.
This is a generalization of
the previous formulation for periodic boundary conditions \cite{AMNS}.
The crucial point was to characterize the
configuration space algebraically.
As a simple example, we
consider 2d U($p$) NC gauge theory with twisted boundary 
conditions,
which correspond to introducing a constant background 
flux specified by the integer $q$.
(We use the notations in ref.\ \cite{Aoki:2008ik} except for switching
the sign of the integer $q$, and hence the sign of $\tilde q$ defined
in (\ref{pp0ptildeqq0qtilde}),
to make some important formulae in this paper look nicer.)


For a gauge-singlet field, the boundary conditions reduce to the 
periodic ones, and the configuration space is given by 
the representation space of the coordinate operators
$\hat Z_\mu = e^{2\pi i \hat x_\mu/L}$
and the shift operators $\hat\Gamma_\mu = e^{\epsilon \hat\del_\mu}$,
which satisfy the algebra
\beqa
\hat Z_\mu \hat Z_\nu &=& e^{-2\pi i \Theta_{\mu\nu}}
\hat Z_\nu \hat Z_\mu \ ,
\label{ncalgZZ}\\
\hat\Gamma_\mu \hat Z_\nu \hat\Gamma_\mu^\dagger
&=&e^{\frac{2\pi i}{N} \delta_{\mu\nu}} \hat Z_\nu \ , 
\label{algGZGZ} \\
\hat\Gamma_\mu \hat\Gamma_\nu &=& 
e^{-i \epsilon ^2 c_{\mu\nu}}
\hat\Gamma_\nu \hat\Gamma_\mu \ .
\label{algGGGG}
\eeqa
Here $\Theta_{\mu\nu}$ and $c_{\mu\nu}$
represent non-commutativity,
and $\epsilon$ serves as the lattice spacing with
\beq
L=\epsilon \, N 
\label{size-original}
\eeq
being the size 
of the torus.

For an adjoint field such as the gauge field,
the theory with twisted boundary conditions 
can be mapped through 
the so-called Morita equivalence
to a U($p_0$) NC gauge theory with periodic boundary
conditions on a dual torus,
where $p_0$ is the greatest common divisor of $p$ and $q$.
For later convenience, 
let us introduce co-prime integers $\tilde p$ and $\tilde q$ by
\beq
p= p_0\, \tilde p \ , \ \ q= p_0\, \tilde q \ ,
\label{pp0ptildeqq0qtilde}
\eeq
and another set of integers $a$ and $b$ by the Diophantine equation
\beq
a \tilde p - b \tilde q = 1 \ .
\label{aibidef}
\eeq
The covariant derivative operator $\hat D^{(0)}_\mu$
for the constant-curvature background field
on the original torus 
is mapped to the derivative operator
$\hat{\del}_\mu ' = \hat D^{(0)}_\mu$
on the dual torus.
Denoting the coordinate operators
and the shift operators on the dual torus by
$\hat Z'_\mu$ and 
$\hat\Gamma'_\mu \equiv e^{\epsilon \hat{\del}_\mu '}
= e^{\epsilon \hat D^{(0)}_\mu}$,
we obtain the algebra
\beqa
\hat Z'_\mu\,\hat Z'_\nu&=&
\e^{-2\pi i\,\Theta'_{\mu\nu}}\,
\hat Z'_\nu\,\hat Z'_\mu \ ,
\label{Zprimealg}\\
\hat\Gamma'_\mu \hat Z'_\nu \hat\Gamma_\mu^{\prime\dagger}
&=&e^{\frac{2\pi i}{n} \delta_{\mu\nu}} \hat Z'_\nu \ , 
\label{algGpZpGpZp} \\
\hat\Gamma'_\mu \hat\Gamma'_\nu 
&=&e^{-i \epsilon^2 (c_{\mu\nu}+f_{\mu\nu})} 
\hat\Gamma'_\nu \hat\Gamma'_\mu \ ,
\label{algGpGpGpGp}
\eeqa
where $\Theta'_{\mu\nu}$ represents the non-commutativity tensor
of the dual torus and
$f_{\mu\nu}$ is the constant background flux specified by the 
integer $q$ as
\beq
f_{12} 
= \frac{-2\pi q}{L^2(p + \Theta q)} \ .
\label{relfluxfq}
\eeq
The non-commutativity tensors are written as
$\Theta_{\mu\nu}=\Theta\varepsilon_{\mu\nu}$
and $\Theta'_{\mu\nu}=\Theta'\varepsilon_{\mu\nu}$
for the original and dual tori, respectively,
where $\varepsilon_{\mu\nu}$ is the anti-symmetric tensor
with $\varepsilon_{12}=1$.
Then, the NC parameter $\Theta'$ and the size
\beq
L'= \epsilon \, n 
\eeq
of the dual torus are determined by those of the original torus as
\beqa
\Theta'&=&\frac{a \Theta + b}{\tilde p +  \tilde q \Theta}\,
\label{Thetaprime}
\ , \\
L'&=&L \,(\tilde p + \Theta \tilde q) \ .
\label{Sigmaprime}
\eeqa
The configuration space of an adjoint field is given by 
the representation space of the operators 
$\hat Z'_\mu$ and  $\hat\Gamma'_\mu$
satisfying the algebra (\ref{Zprimealg})-(\ref{algGpGpGpGp}).

From the above construction of the singlet and adjoint fields,
it follows that
the configuration space of a fundamental field 
is given by the representation space of the operators $\hat Z'_\mu$ and 
$\hat\Gamma'_\mu$ acting from the left,
and the operators $\hat Z_\mu$ and 
$\hat\Gamma_\mu$ acting from the right.

Once the configuration space is characterized algebraically as above,
we can represent the configurations by finite matrices
by finding appropriate 
representations of the coordinate and shift operators.
Since the operators $\hat Z_\mu$ and $\hat\Gamma_\mu$
are defined on the original torus, which is discretized into
an $N \times N$ lattice, it is natural to represent them
by $N \times N$ matrices\footnote{Their explicit form is
given in eq.\ (5.6) of ref.\ \cite{Aoki:2008ik}.
}
from the counting of degrees of freedom.
Then the algebra (\ref{ncalgZZ})-(\ref{algGGGG}) can be satisfied
if the NC parameters are chosen as
\beqa
\Theta &=& \frac{2r}{N} \ ,
\label{ncptheta}\\
\epsilon^2 c_{12}&=& -2 \pi  \frac{s}{N} \ ,
\label{c12}
\eeqa
where the integers $r$ and $s$ satisfy 
the Diophantine equation
\beq
2rs-kN=-1 
\label{diophantinerskN}
\eeq
for some integer $k$.
Similarly, since 
the operators $\hat Z'_\mu$ and $\hat\Gamma'_\mu$
are defined on the dual torus,
which is discretized into an $n \times n$ lattice
and endowed with the U($p_0$) gauge group,
it is natural to represent these operators
by $np_0 \times np_0$ matrices as
\beqa
\hat Z ' _\mu &=& Z_\mu^{(n)} \otimes \id_{p_0}  \ , \\
\hat\Gamma ' _\mu &=& \Gamma_\mu^{(n)} \otimes \id_{p_0}  \ .
\eeqa
The $n \times n$ matrices\footnote{Their explicit form is
given in eq.\ (5.15) of ref.\ \cite{Aoki:2008ik}.
}
$Z_\mu^{(n)}$ and $\Gamma_\mu^{(n)}$
can satisfy the same algebra as
(\ref{Zprimealg})-(\ref{algGpGpGpGp})
if we choose the NC parameters as
\beqa
\Theta' &=& \frac{j}{n} \ ,
\label{ncpthetap}\\
\epsilon^2 (c_{12}+f_{12}) &=& 2\pi \frac{m}{n} \ ,
\label{mnsNf12}
\eeqa
where the integers $j$ and $m$ satisfy 
the Diophantine equation
\beq
mj+nk'=1 \ 
\label{diophantinemjnkp}
\eeq
for some integer $k'$.
The integers $m$ and $n$ of the dual torus 
are determined by those
of the original torus, $N$, $r$, $s$, $k$, with the input of 
the integers $p$ and $q$ as
\beq
m=-s\tilde{p} -k\tilde{q} \ , \ \ \ 
n=N\tilde{p}+2r\tilde{q} \ .
\label{relation-mn-pq}
\eeq
The other integers $j$ and $k'$ of the dual torus
can be determined by eq.\ (\ref{diophantinemjnkp}).


The action for the gauge field is given by 
the twisted Eguchi-Kawai model \cite{EK,GAO}
\beq
S_{\rm TEK} = -n \beta' \, \sum_{\mu \ne \nu} 
{\cal  Z'}_{\nu\mu}
\tr ~\Bigl(V_\mu\,V_\nu\,V_\mu^\dag\,V_\nu^\dag\Bigr) 
 + 2 \beta' n^2 p_0 \ ,
\label{TEK-action}
\eeq
where
\beq
V_\mu = e ^{\epsilon \hat{D}_\mu} 
\label{defV}
\eeq
are U($np_0$) matrices
with $\hat{D}_\mu$ being the covariant derivative operator
for the full gauge field including both the background and fluctuations.
The $Z_n$ factor ${\cal  Z'}_{\nu\mu}=({\cal  Z'}_{\mu\nu})^{*}$ represents
the twist, which is given by
\beq
{\cal  Z'}_{12} = \exp{\left(-2\pi i \frac{m}{n}\right)} 
\ .
\label{twistZ12}
\eeq
The coefficient $\beta'$ 
can be interpreted as the lattice coupling constant
of the dual theory,
which is related to that of the original theory $\beta$ as
\beq
\beta = \frac{1}{\tilde{p}} \left(\frac{n}{N} \right)^2 \beta' \ .
\label{rel_beta_betaprime}
\eeq
The action (\ref{TEK-action}) takes the minimum value and vanishes
when $V_\mu = \hat{\Gamma}'_\mu \equiv 
e^{\epsilon \hat D^{(0)}_\mu}$,
which corresponds to the background gauge field.
One can also show that the action (\ref{TEK-action})
has the correct continuum limit classically.

Actions for fundamental matters can be given by using 
the covariant forward and backward difference operators 
$\nabla_\mu$, $\nabla_\mu^*$ defined by
\beqa
\nabla_\mu \Psi&=&
\frac{1}{\epsilon}\left(V_\mu \, \Psi \, \hat{\Gamma}_\mu ^{\dag}
- \Psi  \right) \ , \n
\nabla_\mu^* \Psi &=&
\frac{1}{\epsilon}\left(\Psi - V_\mu ^{\dagger} \,
\Psi \,  \hat{\Gamma}_\mu \right)  \  .
\label{def-cov-shift}
\eeqa
Here  $V_\mu$ is the U$(n p_0)$ matrix
introduced by (\ref{defV}),
and $\hat{\Gamma}_\mu=\e^{\epsilon \hat\partial_\mu}$ 
is the shift operator represented by an $N \times N$ matrix
satisfying (\ref{algGGGG}) and (\ref{c12}).
The fundamental matter field $\Psi$ is represented by 
$np_0 \times N$ rectangular matrices.
One can define an overlap 
Dirac operator \cite{Neuberger} as\footnote{The overlap 
Dirac operator was
introduced on a periodic NC torus in ref.\ \cite{Nishimura:2001dq},
and the correct form of the axial anomaly has been reproduced
in the continuum limit \cite{isonagao}.
A prescription to define an analog of the overlap Dirac operator
and its index (\ref{def_nu})
on general NC manifolds including the fuzzy sphere
has been proposed in ref.\ \cite{AIN2}.
}
\beq
D=\frac{1}{\epsilon} (1 - \gamma_5 \hat{\gamma}_5) \ ,
\label{def-GW-Dirac}
\eeq
where $\gamma_5$ 
is the ordinary chirality operator and $\hat{\gamma}_5$
is the modified one defined by
\beqa
\hat\gamma_5 &=& \frac{H}{\sqrt{H^2}} \ , \\
H &=& \gamma_5 \left(1- \epsilon D_{\rm W}\right) 
\label{H-def}
\eeqa
in terms of the Wilson-Dirac operator 
\beq
D_{\rm W}=\frac{1}{2}\sum_{\mu=1}^2
\left\{\gamma_\mu\left(\nabla_\mu^* 
+\nabla_\mu \right) - \epsilon  \nabla_\mu^* \nabla_\mu \right\} \ .
\label{def-Wilson-Dirac}
\eeq
The Dirac operator (\ref{def-GW-Dirac}) satisfies
the Ginsparg-Wilson relation \cite{GinspargWilson}
\beq
\gamma_5 D + D \hat\gamma_5 =0 \ ,
\eeq
which guarantees the exact 
chiral symmetry \cite{Luscher}.
Thanks to the index theorem \cite{Hasenfratzindex},
one can 
classify gauge configurations into topological sectors
using the index of $D$ given by
\beq
\nu = \frac{1}{2} \,  \CTr \left(\gamma_5+\hat\gamma_5 \right)
= \frac{1}{2} \, \CTr \, \hat\gamma_5 \ ,
\label{def_nu}
\eeq
where the trace $\CTr$ is taken
in the configuration space of the matter field.

The topological charge can also be defined as\footnote{This
reduces to the one in ref.\ \cite{Aoki:2006sb} for $q=0$.
An analogous definition was also used
in ref.\ \cite{Griguolo:2003kq}.
Their topological charge $Q_{\rm GS}$ is related to ours
by $Q = -Q_{\rm GS}/2 \pi$.
}
\begin{align}
 Q = \frac{1}{4 \pi i} N \, \sum_{\mu \ne \nu} \epsilon_{\mu\nu}
{\cal  Z}_{\nu\mu}
\tr ~\Bigl(V_\mu\,V_\nu\,V_\mu^\dag\,V_\nu^\dag\Bigr) 
\label{naivetopcharge} 
\end{align}
with the $Z_{N}$ factor 
${\cal  Z}_{\nu\mu}=({\cal  Z}_{\mu\nu})^{*}$ given by
\begin{align}
{\cal  Z}_{12} = \exp{\left( 2\pi i \frac{s}{N} \right)} \ .
\end{align}
For $V_\mu = \hat{\Gamma}'_\mu$, which gives the
minimum of the gauge action (\ref{TEK-action}),
the topological charge becomes $Q=q$ as expected.
One can also show that $Q$ has the correct continuum limit
classically.
Note, however, that $Q$ does not take integer values 
for generic configurations unlike the index $\nu$.

\section{Monte Carlo results}
\label{sec:index}

In this section we present our results for 
the probability distribution
of the index (\ref{def_nu}) and 
the topological charge (\ref{naivetopcharge})
obtained by simulating the model (\ref{TEK-action}).
Details of the simulation are
described in ref.\ \cite{2dU1}.

First we comment on the parameters we have chosen.
For simplicity, we consider U(1) gauge group $p=1$,
which also implies $p_0=1$ and hence $\tilde p = 1$,
$\tilde q = q$.
As for the integers $r$ and $k$
appearing in eq.\ (\ref{diophantinerskN}),
we choose $r=-1$, $k=-1$ (and hence $s=\frac{N+1}{2}$)
following essentially 
the choice in the previous works 
\cite{2dU1,Aoki:2006sb,Aoki:2006zi,Aoki:2008ik}.
This implies, in particular, 
that the NC parameter (\ref{ncptheta})
is given by $\Theta = -2/N$, where $N$ represents
the size of the torus (\ref{size-original})
in units of the lattice spacing.
Note that the size $n$ of the matrices $V_\mu$
and 
the integer $m$, which labels the twist (\ref{twistZ12})
in the gauge action (\ref{TEK-action}),
are given by $n=N-2q$ and $m=-(n+1)/2$, respectively,
due to (\ref{relation-mn-pq}).

We perform Monte Carlo simulations for various values of $q$.
We measure the index $\nu$
and the topological charge $Q$
for each configuration generated by Monte Carlo
simulation, 
and obtain the probability distribution $P(\nu)$ and $P(Q)$. 
%
%

\begin{figure}[H]
\begin{center}
\begin{minipage}{.45\linewidth}
\includegraphics[width=0.80 \linewidth]{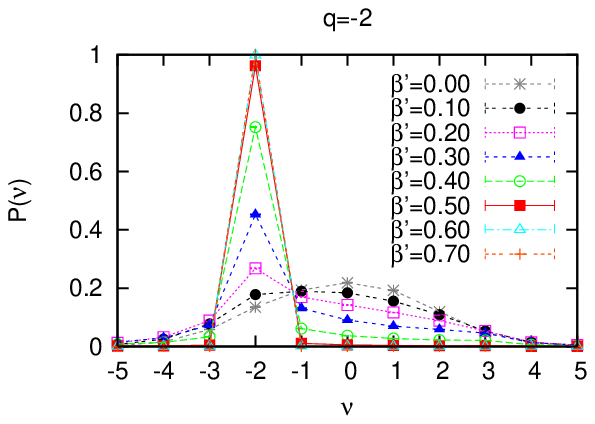}
  \end{minipage}
  \hspace{1.0pc}
\begin{minipage}{.45\linewidth}
\includegraphics[width=0.80 \linewidth]{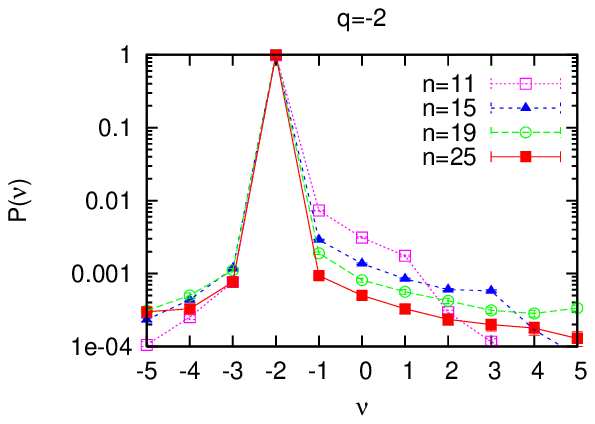}
  \end{minipage}

\begin{minipage}{.45\linewidth}
\includegraphics[width=0.80 \linewidth]{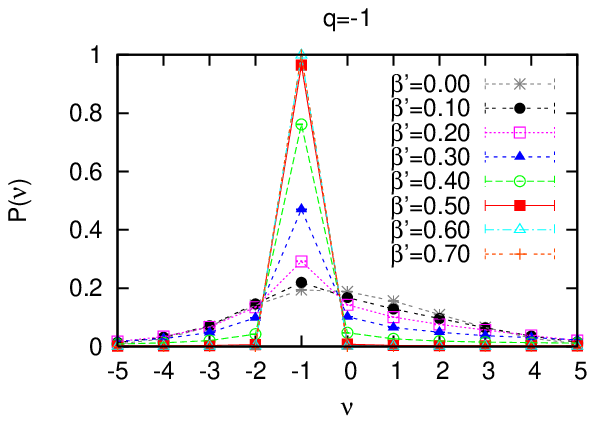}
  \end{minipage}
  \hspace{1.0pc}
\begin{minipage}{.45\linewidth}
\includegraphics[width=0.80 \linewidth]{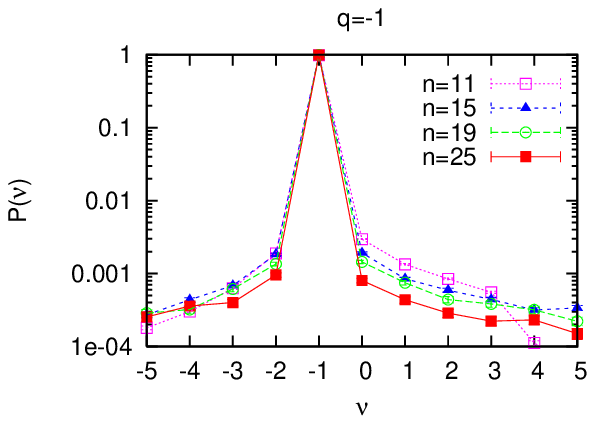}
  \end{minipage}

\begin{minipage}{.45\linewidth}
\includegraphics[width=0.80 \linewidth]{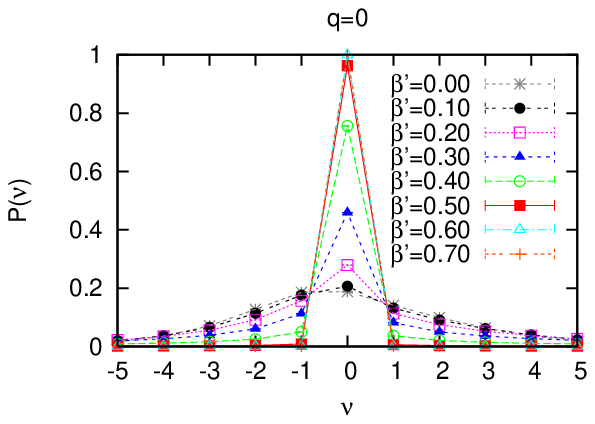}
  \end{minipage}
  \hspace{1.0pc}
\begin{minipage}{.45\linewidth}
\includegraphics[width=0.80 \linewidth]{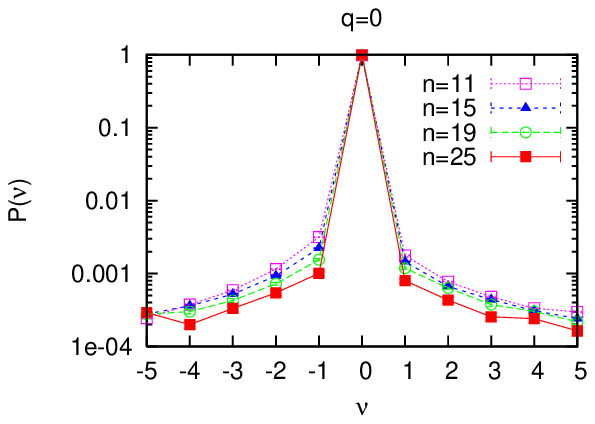}
  \end{minipage}


\begin{minipage}{.45\linewidth}
\includegraphics[width=0.80 \linewidth]{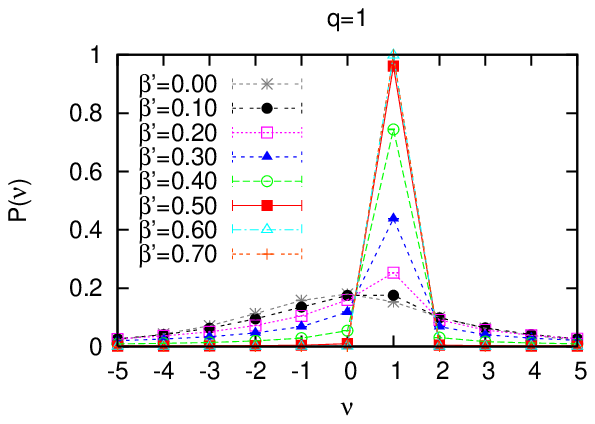}
  \end{minipage}
  \hspace{1.0pc}
\begin{minipage}{.45\linewidth}
\includegraphics[width=0.80 \linewidth]{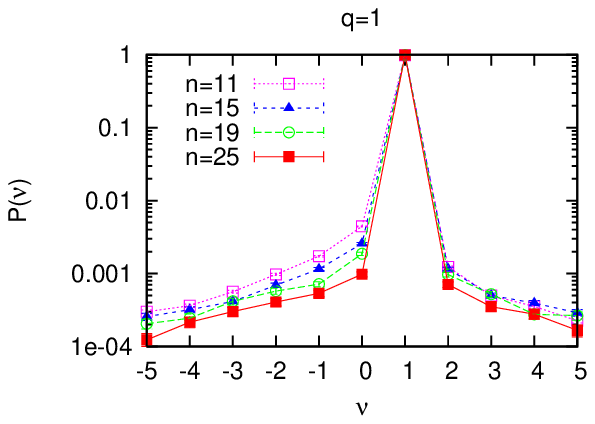}
  \end{minipage}

\begin{minipage}{.45\linewidth}
\includegraphics[width=0.80 \linewidth]{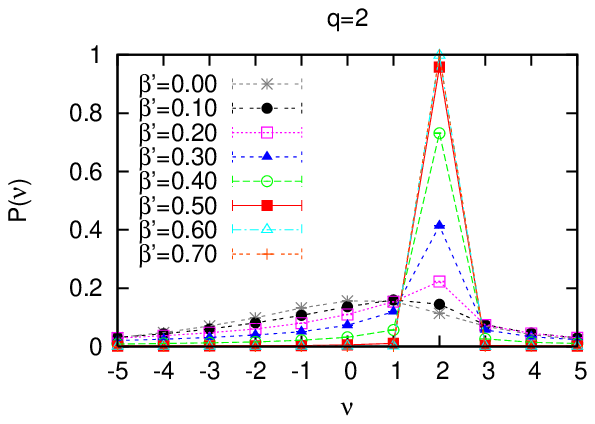}
  \end{minipage}
  \hspace{1.0pc}
\begin{minipage}{.45\linewidth}
\includegraphics[width=0.80 \linewidth]{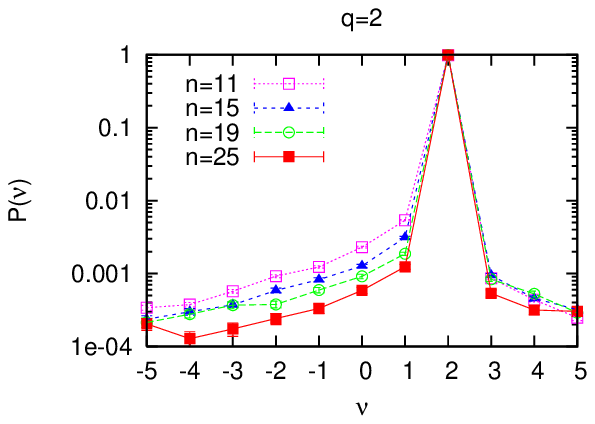}
  \end{minipage}

\end{center}
\caption{The probability distribution of the index $\nu$ 
is plotted for various $\beta '$ at $n=15$ (left) and 
for various $n$ at $\beta '=0.55$ (right).
{}From the top to the bottom, we present the results for
$q=-2,-1,0,1,2$.
In the plots on the right, the probability is plotted in
the log scale to make the distribution at $\nu \neq q$ visible.
}
\label{distrib-N-beta}
  \end{figure}

Let us present our results for the index 
$\nu$ 
given by (\ref{def_nu}).
In fig.\ \ref{distrib-N-beta} we plot the distribution
of the index for
$q=-2,-1,0,1,2$.
(The results for $q=0$ are already given in ref.\ \cite{Aoki:2006zi}.)
On the left we show
how the probability distribution of $\nu$ changes 
as we increase $\beta'$ for $n=15$.
(We assume the normalization
$\sum_\nu P(\nu) = 1$.)
We find that the probability for $\nu \neq q$ decreases rapidly,
and the probability for $\nu = q$ approaches unity.
On the right we plot 
the probability distribution $P(\nu)$
for various $n$ at $\beta' = 0.55$.
(Note that the value of $\beta'$ we have chosen
lies in the region above the 
critical point $\beta' = \beta_{\rm cr}\equiv 1/2$
of the Gross-Witten phase transition \cite{GW},
which is relevant for the continuum limit.)
We find that the distribution approaches
the Kronecker delta $\delta_{\nu , q}$
not only for increasing $\beta'$ but also for increasing $n$.

\begin{figure}[H]
\begin{center}

\begin{minipage}{.45\linewidth}
\includegraphics[width=0.88 \linewidth]{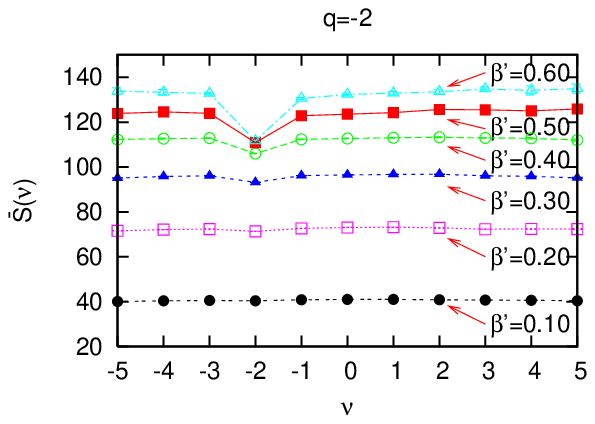}
  \end{minipage}
  \hspace{1.0pc}
\begin{minipage}{.45\linewidth}
\includegraphics[width=0.88 \linewidth]{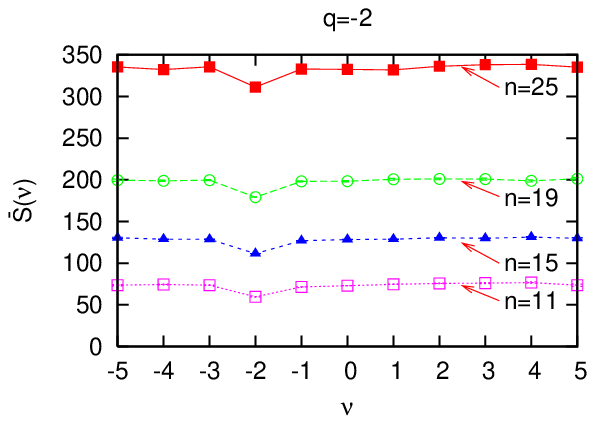}
  \end{minipage}

\begin{minipage}{.45\linewidth}
\includegraphics[width=0.88 \linewidth]{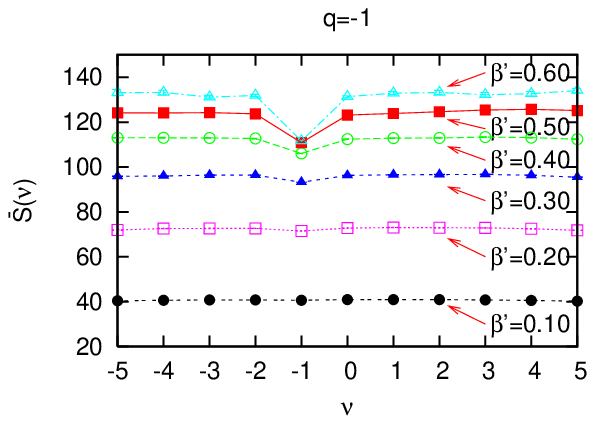}
  \end{minipage}
  \hspace{1.0pc}
\begin{minipage}{.45\linewidth}
\includegraphics[width=0.88 \linewidth]{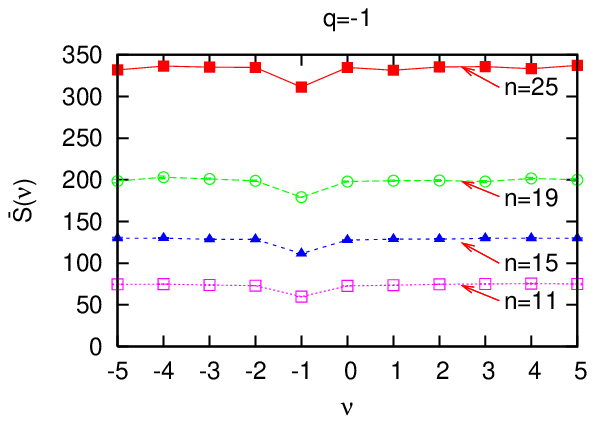}
  \end{minipage}

\begin{minipage}{.45\linewidth}
\includegraphics[width=0.88 \linewidth]{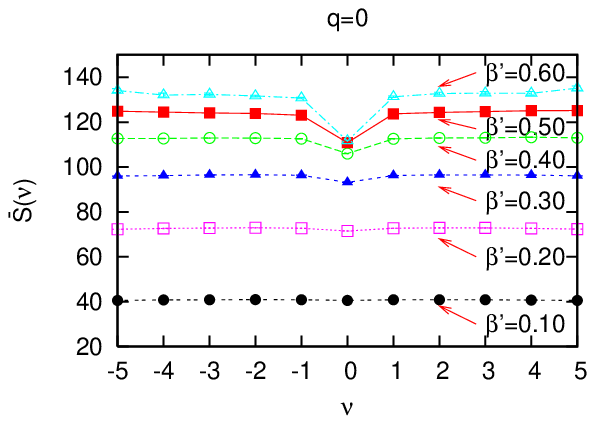}
  \end{minipage}
  \hspace{1.0pc}
\begin{minipage}{.45\linewidth}
\includegraphics[width=0.88 \linewidth]{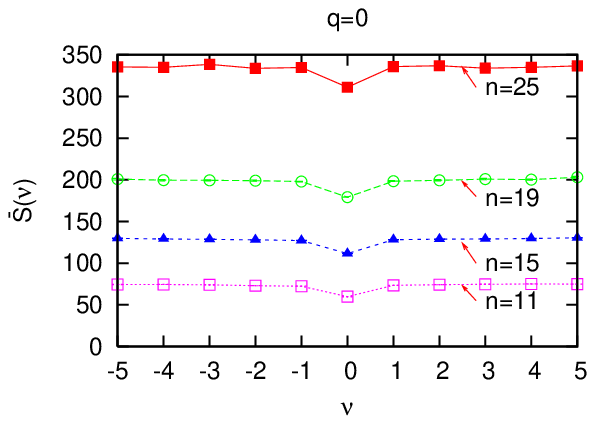}
  \end{minipage}


\begin{minipage}{.45\linewidth}
\includegraphics[width=0.88 \linewidth]{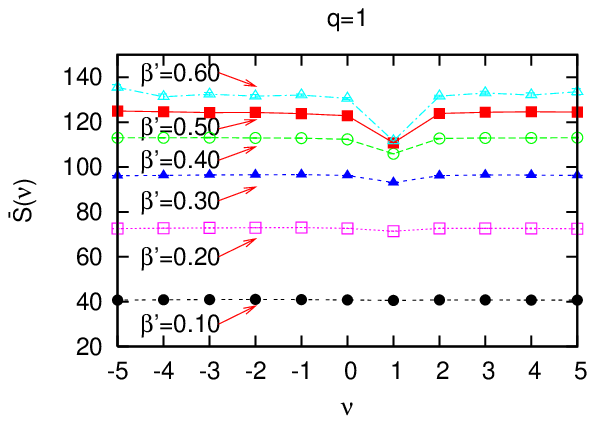}
  \end{minipage}
  \hspace{1.0pc}
\begin{minipage}{.45\linewidth}
\includegraphics[width=0.88 \linewidth]{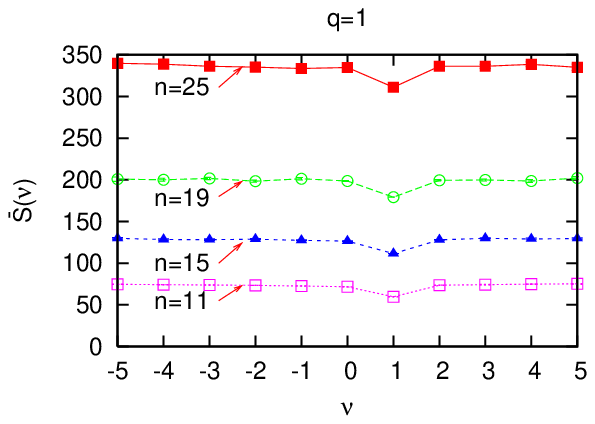}
  \end{minipage}

\begin{minipage}{.45\linewidth}
\includegraphics[width=0.88 \linewidth]{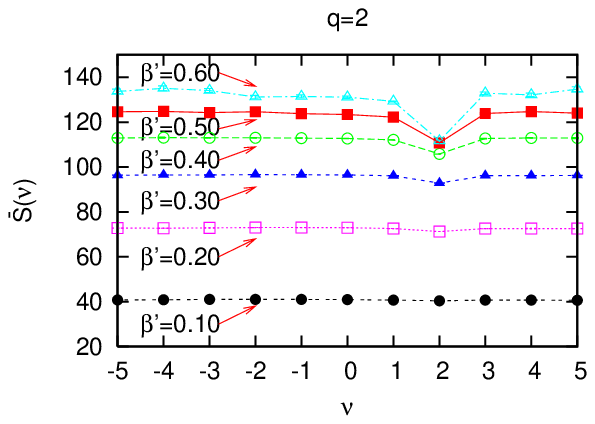}
  \end{minipage}
  \hspace{1.0pc}
\begin{minipage}{.45\linewidth}
\includegraphics[width=0.88 \linewidth]{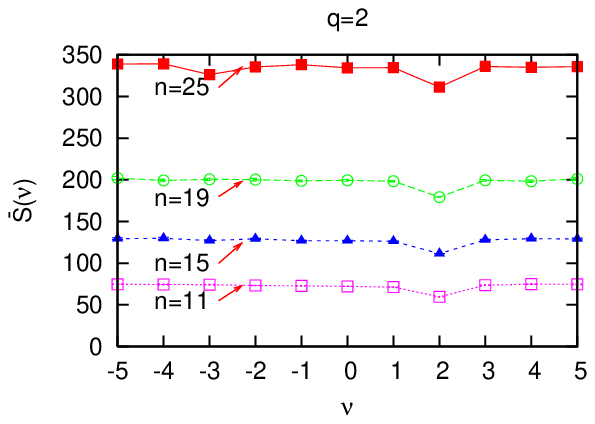}
  \end{minipage}

\end{center}
\caption{The average value of the action
is plotted 
against the index $\nu$ for various $\beta'$ at $n=15$ (left)
and for various $n$ at $\beta'=0.55$ (right).
{}From the top to the bottom, we present the results for
$q=-2,-1,0,1,2$.
}
\label{actionN15}
  \end{figure}

\begin{figure}[H]
\begin{center}
\includegraphics[width=0.40 \linewidth]{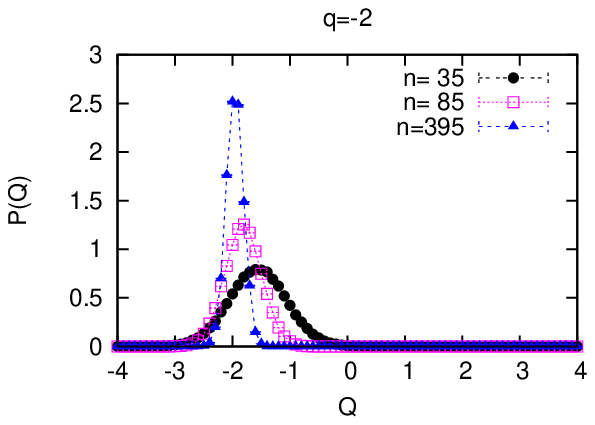}
\end{center}
\vspace{-1cm}
\begin{center}
\includegraphics[width=0.40 \linewidth]{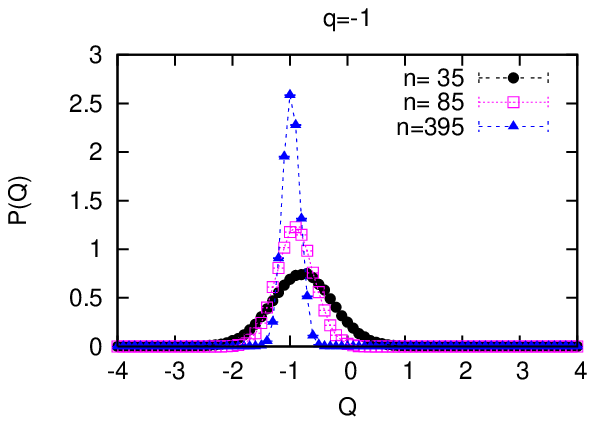}
\end{center}
\vspace{-1cm}
\begin{center}
\includegraphics[width=0.40 \linewidth]{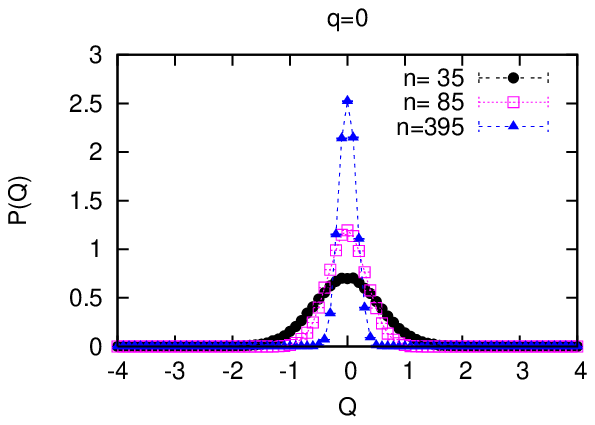}
\end{center}
\vspace{-1cm}
\begin{center}
\includegraphics[width=0.40 \linewidth]{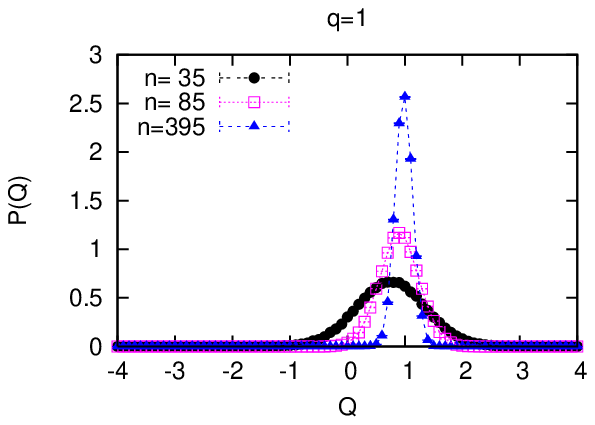}
\end{center}
\vspace{-1cm}
\begin{center}
\includegraphics[width=0.40 \linewidth]{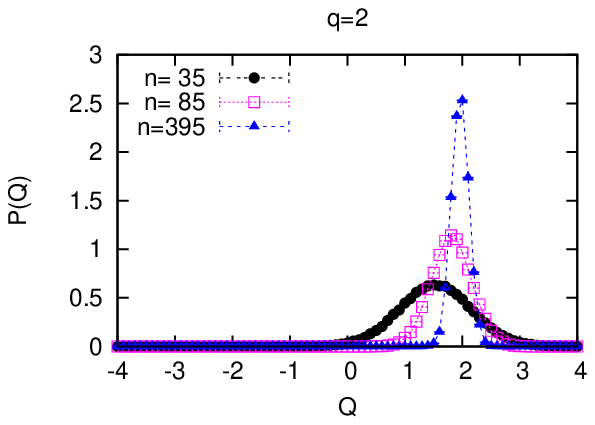}
\end{center}
\caption{The probability distribution of the topological charge
$Q$ is plotted for $q=-2,-1,0,1,2$ with $n=35,85,395$
and fixed $n/\beta'=32$.
}
\label{distrib-Q-DSL}
  \end{figure}

\FIGURE{
    \epsfig{file=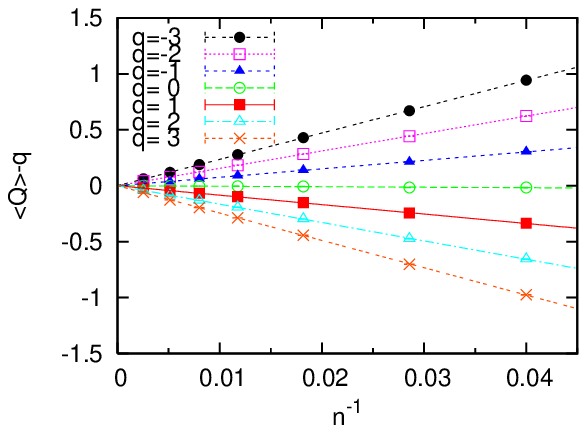,%
width=7.4cm}
    \epsfig{file=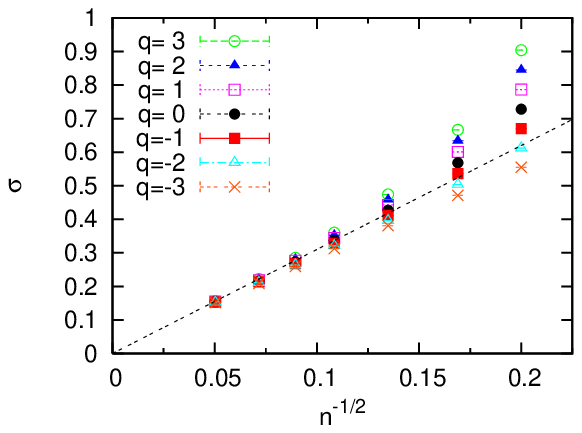,%
width=7.4cm}
\caption{(Left) The discrepancy of the vacuum expectation value of
the topological charge $Q$
from $q$ is plotted against $1/n$ .
(Right) The width of the probability distribution of $Q$ 
is plotted against $1/\sqrt{n}$.
}
\label{vevQ-sigma}
}

In order to take the continuum limit,
we have to send $n$ and $\beta'$ to infinity
simultaneously fixing the ratio
$n/\beta'$ \cite{2dU1}.\footnote{The ``lattice spacing'' $\epsilon$
has to be sent to zero in such a way that the dimensionful coupling
constant $g^2 = \frac{1}{\epsilon ^2 \beta}$ is fixed.
We also require the dimensionful non-commutativity
parameter $\Theta L^2 \sim N \epsilon^2 $ 
to be finite. These two conditions imply
fixing the ratio $N/\beta \sim n/\beta'$ 
in the continuum limit.
The size of the torus 
$L = N \epsilon$
diverges as 
$\sqrt{N}\sim \sqrt{n}$ in that limit. 
}
It is clear from the above results that
the distribution $P(\nu)$ approaches $\delta_{\nu , q}$
very rapidly in that limit.
This conclusion 
provides a physical interpretation
to the results of the instanton calculus
in the continuum theory \cite{Paniak:2002fi},
where the partition function has been written
as a sum over all the instanton configurations
with a certain constraint
related to the magnetic flux $q$.

In the commutative case,
analytical and numerical studies show that 
the distribution of the topological charge
is gaussian with a finite width, and the 
width diverges in the infinite-volume limit
 \cite{Hassan:1995dn,GHL}.
Thus the situation in the NC case differs
drastically from the commutative case.
This can be interpreted as a 
kind of smoothing effects of 
NC geometry due to the existence of all higher derivative
terms in the star product as we discussed in the Introduction.
To substantiate this argument,
we plot in fig. \ref{actionN15}
the average value of the action $\bar{S}(\nu)$ in 
each topological sector.
We find that the result is almost independent of $\nu$
except for $\nu = q$, where we indeed observe a dip.

Next we present our results for 
the topological charge $Q$ defined by (\ref{naivetopcharge}).
In fig.\ \ref{distrib-Q-DSL}
the probability distribution of 
$Q$ is plotted for $q=-2,-1,0,1,2$,
showing how it changes as we take the continuum limit,\footnote{In
ref.\ \cite{Frisch:2007zz},
growing of the peak is observed
as one increases $\beta'$ with fixed $n$
for $q=0$.}
{\it i.e.},
sending $n$ and $\beta'$ to infinity 
with fixed $n/\beta'$.
Here we fix it to $n/\beta'=32$ following ref.\ \cite{2dU1}.
(We assume the normalization $\int dQ \ P(Q) =1$.)
We find that the distribution is a gaussian
peaked slightly away from the integer $q$.
From fig.\ \ref{vevQ-sigma},
however, we find in the continuum limit that the deviation 
$\langle Q \rangle - q$
vanishes as $1/n$,
and
the width of the distribution vanishes
as $1/\sqrt{n}$.
Thus we have confirmed that the distribution of the topological
charge becomes a delta function peaked at $q$.
This is consistent with the index 
theorem \cite{Atiyah:1971rm,Kim:2002qm}, 
which asserts
that the index $\nu$ agrees with the topological charge $Q$
for configurations that survive in the continuum limit.

Let us recall that the definition of the 
topological charge we used is given by eq.\ (\ref{naivetopcharge}),
which may be viewed as a naive discretization 
of the corresponding expression in the continuum. 
In the commutative space, one can use
the geometrical construction \cite{Luscher:1981zq}
to define an integer topological charge 
for any lattice configuration.
It takes a particularly simple form in the 2d U(1) case, 
which is used in ref.\ \cite{Hassan:1995dn, GHL}.
In 4d theories, on the other hand, 
one usually uses a naively defined topological charge,
but one can obtain a distribution peaked at integer values
by using some techniques like cooling or renormalization.
(See ref.\ \cite{Vicari:2008jw}, for instance.)
Note that here we are able to obtain 
a distribution peaked at an integer value
even with the naive definition of the topological charge
without any techniques.
This might also be due to smoothing effects of NC geometry.
Let us also mention that the computational effort for calculating
$Q$ is of order O($n^3$), whereas that for calculating the index $\nu$
is of order O($n^6$). Therefore, $Q$ might be of some use
when one tries to extend the present work to higher dimensions.

\section{Summary and discussions}
\label{sec:summary}

In this paper, we have studied the probability distribution of 
the index $\nu$ and the topological charge $Q$
in the finite-matrix formulation of 2d U(1) NC gauge theory
with various boundary conditions.
Our results suggest
that a single topological sector,
which is dictated by 
the boundary condition specified by the integer $q$,
dominates
in the continuum limit.\footnote{In the literature,
the possibility of summing over boundary conditions is sometimes discussed.
In this case, the relative weight
should be fixed by some underlying principle.
For example, in the random matrix theory for QCD, the relative
weight is fixed by the guiding principle that one should
recover QCD \cite{Shuryak:1992pi}.
Alternatively, one can introduce the Higgs sector in the
NC gauge theory \cite{Aoki:2006wv}
(See \cite{Steinacker:2003sd} for a related work.).
In this case, the winding number of the Higgs field plays
the role of the twisted boundary condition, 
and all the topological sectors appear from a single theory.
%
%
}
This is in sharp contrast to the situation in 
the ordinary lattice gauge theory, where
all the topological sectors appear from a single theory.
This striking difference can be interpreted as a kind of
smoothing effects of NC geometry.

Let us discuss possible 
implications of our results to the real world.
Considering that
our space-(time) naturally becomes non-commutative,
for instance, in matrix model formulations 
\cite{Banks:1996vh,Ishibashi:1996xs}
of string theory, we may expect that NC geometry is realized
in nature in various ways, which we discuss in what follows separately.

First we consider the possibility that NC geometry is realized 
in the extra dimensions \cite{Aschieri:2003vy,AIMN}.
There is a well-known scenario in string theory 
that the number of generations, for instance,
is determined by the index
in the extra dimensions.
However, its distribution
may have
a non-vanishing width in general
due to dynamical fluctuations of 
the gauge fields
in the extra dimensions, which causes a serious problem
in this scenario.
If the extra-dimensional space is non-commutative, 
the index takes a specific value once the 
boundary condition is fixed somehow,
even if the dynamical fluctuations are taken into account.

We can also consider that our 4d space-time has
certain non-commutativity.
In this case, non-locality may cause some problems 
such as the violation of causality, renormalizability
and the CPT theorem.
It also induces the UV/IR mixing effects, 
which make it difficult to think of phenomenologically viable models.
From the viewpoint of string theory, 
it is a big challenge to understand how one can get rid of
all these problems arising from non-locality.
(See ref.\ \cite{Balachandran:2006pi} for a different
way to introduce NC geometry without inducing the UV/IR mixing effects.)
Let us assume here that they are resolved somehow, and
discuss what would be other consequences of the NC geometry.
In particular,
we speculate on how the new properties we found in the 
present paper may play roles in various problems 
related to topological aspects of gauge theories.

Let us consider the baryon number asymmetry of the universe. 
In the electroweak theory,
the baryon number conservation is violated by the chiral anomaly
through instanton effects.
In order to generate net baryon number asymmetry,
one usually introduces some bias 
caused, for instance, by
a bubble wall sweeping over the universe 
during the electroweak phase transition.
Our result suggests that
NC geometry with 
a twisted boundary condition may provide 
a simple and direct source
for the net asymmetry.

We also speculate that the strong CP problem 
may be solved by NC geometry. 
From the experimental bound on the electric dipole moment,
the parameter $\bar\theta$ of the QCD vacuum 
(including the phase of the fermion mass)
must be less than $10^{-9}$,
which needs some explanation.
If the space-time is given by a NC geometry,
a single topological sector with a specific value of $Q$ 
dominates, and the $e^{i \bar\theta Q}$
simply factors out of the path integral.
Hence the physics do not depend on the parameter $\bar\theta$.
This explanation is reminiscent of the 
scenario \cite{Samuel:1991cm} that the long-range interaction 
between instantons and anti-instantons 
washes out the local topological structure, which
makes the topology of the gauge field 
determined only by the boundary condition.

We hope that some of the above speculations can be pursued further
by investigating the relevance of matrix model formulations of 
string theory to our world, and by studying NC gauge theories in four
dimensions by Monte Carlo simulations \cite{4dU1}.

\acknowledgments

We thank
Satoshi Iso, Hikaru Kawai and Hiroshi Yoneyama 
for valuable discussions.

\end{document}